\documentclass[pra,10pt,showpacs,amsmath,twocolumn,floatfix,eqsecnum,superscriptaddress]{revtex4}
\usepackage{amsmath}
\usepackage{amsfonts}
\usepackage{graphicx}

\newcommand{\parti}[2]{\frac{\partial #1}{\partial #2}}

\newcommand{\diff}[2]{\frac{d #1}{d #2}}
\newcommand{\intall}{\int_{-\infty}^{\infty}}
\newcommand{\ket}[1]{|#1\rangle}
\newcommand{\bra}[1]{\langle#1|}

\newcommand{\avg}[1]{\langle#1\rangle}

\newcommand{\sinc}{\operatorname{sinc}}

\newcommand{\sg}{\hat{\mathcal{E}}}
\newcommand{\bs}[1]{\boldsymbol{#1}}
\newcommand{\abs}[1]{\left|#1\right|}
\newcommand{\bk}[1]{\left(#1\right)}
\newcommand{\Bk}[1]{\left[#1\right]}

\newcommand{\trace}{\operatorname{Tr}}
\newcommand{\est}[1]{\widetilde{#1}}
\begin{document}
\title{Quantum theory of optical temporal phase and instantaneous
  frequency}

\author{Mankei Tsang}

\email{mankei@mit.edu}

\author{Jeffrey H.\ Shapiro}

\affiliation{Research Laboratory of Electronics, Massachusetts
  Institute of Technology, Cambridge, Massachusetts 02139, USA}

\author{Seth Lloyd}

\affiliation{Research Laboratory of Electronics,
Massachusetts Institute of Technology, Cambridge, Massachusetts
02139, USA}

\affiliation{Department of Mechanical Engineering,
Massachusetts Institute of Technology, Cambridge, Massachusetts
02139, USA}

\date{\today}

\begin{abstract}
  We propose a general quantum theory of optical phase and
  instantaneous frequency in the time domain for slowly varying
  optical signals.  Guided by classical estimation theory, we design
  homodyne phase-locked loops that enable quantum-limited measurements
  of temporal phase and instantaneous frequency. Standard and
  Heisenberg quantum limits to such measurements are then derived. For
  optical sensing applications, we propose multipass and Fabry-P\'erot
  position and velocity sensors that take advantage of the
  signal-to-noise-ratio enhancement effect of wideband angle
  modulation without requiring nonclassical light. We also generalize
  our theory to three spatial dimensions for nonrelativistic bosons
  and define an Hermitian fluid velocity operator, which provides a
  theoretical underpinning to the current-algebra approach of quantum
  hydrodynamics.
\end{abstract}

\maketitle

\section{Introduction}
Frequency modulation (FM) radio sounds better than amplitude
modulation (AM) radio, because by encoding a message in the
instantaneous frequency, defined as the rate of change of the signal
phase, FM uses more transmission bandwidth to improve the
signal-to-noise ratio (SNR) \cite{viterbi,vantrees,couch}. In optical
communications, FM and phase modulation (PM) techniques have also
received significant attention and can offer higher SNRs in photonic
links than standard intensity modulation \cite{kalman}. Quantum
statistics is expected to play a major role in the performance of
coherent optical communication systems \cite{yamamoto,haus,yuen_book},
but although a treatment of digital FM quantum noise has been briefly
mentioned by Yuen \cite{yuen_book}, a more fundamental quantum theory
of temporal phase and instantaneous frequency is not yet available. In
optical sensing, on the other hand, a considerable amount of research
has been devoted to the study of ultimate quantum limits to phase
\cite{caves_prd,shapiro,glm,higgins} and velocity \cite{braginsky}
measurements. A quantum description of temporal phase and
instantaneous frequency is again needed to deal with any rapid change
in the measured parameters. For example, in laser Doppler velocimetry,
which has been widely used in fluid dynamics \cite{bachalo} and blood
flow diagnostics \cite{oberg}, the velocity of the interrogated sample
actually shifts the instantaneous frequency of the reflected optical
signal and not the average frequency directly, so the quantum limits
to velocity measurements derived in Ref.~\cite{braginsky} are no
longer accurate when the velocity changes rapidly.

A rigorous quantum description of the optical phase is, unfortunately,
not trivial even for one optical mode, and has been a subject of
long-running debate
\cite{shapiro,susskind,pegg,noh,leonhardt,mandel,helstrom,holevo}
since the issue was first raised by Dirac at the birth of quantum
electrodynamics \cite{dirac}. The difficulty of defining a phase
operator is shared by the superfluid and Bose-Einstein condensate
community \cite{anderson}, where the fluid velocity in Landau's
formulation of quantum hydrodynamics \cite{landau} is often related to
the phase gradient of an order parameter. Beyond the Bogoliubov
approximation, while Landau has proposed a fluid velocity operator
\cite{landau}, many have argued against its existence \cite{london}.

In this paper, we show how the temporal phase and the instantaneous
frequency of a slowly varying optical signal can be treated
consistently in the quantum regime. By limiting the bandwidth of the
optical Hilbert space, we discretize the continuous time domain into
orthogonal ``wave-packet'' modes via the sampling theorem. This
crucial step allows us to apply previous studies of discrete-mode
quantum phase to the time domain. To enable quantum-limited
temporal-phase measurements in practice, we use estimation theory to
design homodyne phase-locked-loop measurement schemes, similar to
adaptive phase measurements \cite{wiseman,berry_limit,berry}. Standard
and Heisenberg quantum limits to the accuracy of temporal-phase and
instantaneous-frequency measurements for coherent states and squeezed
states are then derived. For sensing applications, we propose
multipass and Fabry-P\'erot position and velocity sensors that take
advantage of the SNR enhancement effect of wideband angle modulation.
Finally, we generalize our time-domain formalism for optics to three
spatial dimensions for nonrelativistic bosons and define an Hermitian
fluid velocity operator, which provides a theoretical underpinning for
the current-algebra approach to quantum hydrodynamics
\cite{landau,london,anderson,dashen}.

\section{\label{quantize}Quantization of Bandlimited Optical Fields}
To describe a typical free-space or fiber optical experiment,
propagating modes are usually quantized in terms of the frequency,
transverse momentum, and polarization degrees of freedom
\cite{haus,yuen,shapiro_qso}. We shall consider a plane wave of a
certain polarization in free space or one fiber mode at an observation
plane along the optical axis, and allow only the frequency degree of
freedom for simplicity. In the slowly varying envelope regime, one can
regard the optical fields as a one-dimensional nonrelativistic
many-boson system and quantize accordingly
\cite{haus,yuen,shapiro_qso,fetter}.  The positive-frequency
electric-field operator can then be approximated as
\begin{align}
\hat{E}^{(+)}(t) \propto \hat{A}(t)\exp(-i2\pi f_0 t),
\end{align}
where $\hat{A}(t)$ is the envelope annihilation operator and $f_0$ is
the carrier optical frequency.  We shall make the physically
reasonable assumption that our ability to create, manipulate,
transmit, and measure the optical fields is limited to a certain
bandwidth $B$ near the carrier optical frequency $f_0$. We can
then express $\hat{A}(t)$ as the bandlimited Fourier transform
of the frequency-domain annihilation operator $\hat{a}(f)$,
\begin{align}
\hat{A}(t) &= \int_{-B/2}^{B/2} df \hat{a}(f)\exp(-i2\pi f t).
\end{align}
In other words, we only need the set of operators $\hat{a}(f)$ and
$\hat{a}^\dagger(f)$ for which $f$ falls within the band ($|f|<B/2$)
to describe our experiments. To satisfy the slowly varying envelope
approximation, we require $B/2 \ll f_0$. The commutation relation
between $\hat{a}(f)$ and the corresponding creation operator
$\hat{a}^\dagger(f)$ is
\begin{align}
[\hat{a}(f),\hat{a}^\dagger(f')]=\delta(f-f'),
\end{align}
which leads to the following time-domain commutation relation,
\begin{align}
[\hat{A}(t),\hat{A}^\dagger(t')] &= B\sinc B(t-t'),
\label{commutator}
\end{align}
where the $\sinc$ function is defined as
\begin{align}
\sinc(x) &\equiv \frac{\sin(\pi x)}{\pi x}.
\end{align}

\begin{figure}[htbp]
\includegraphics[width=0.4\textwidth]{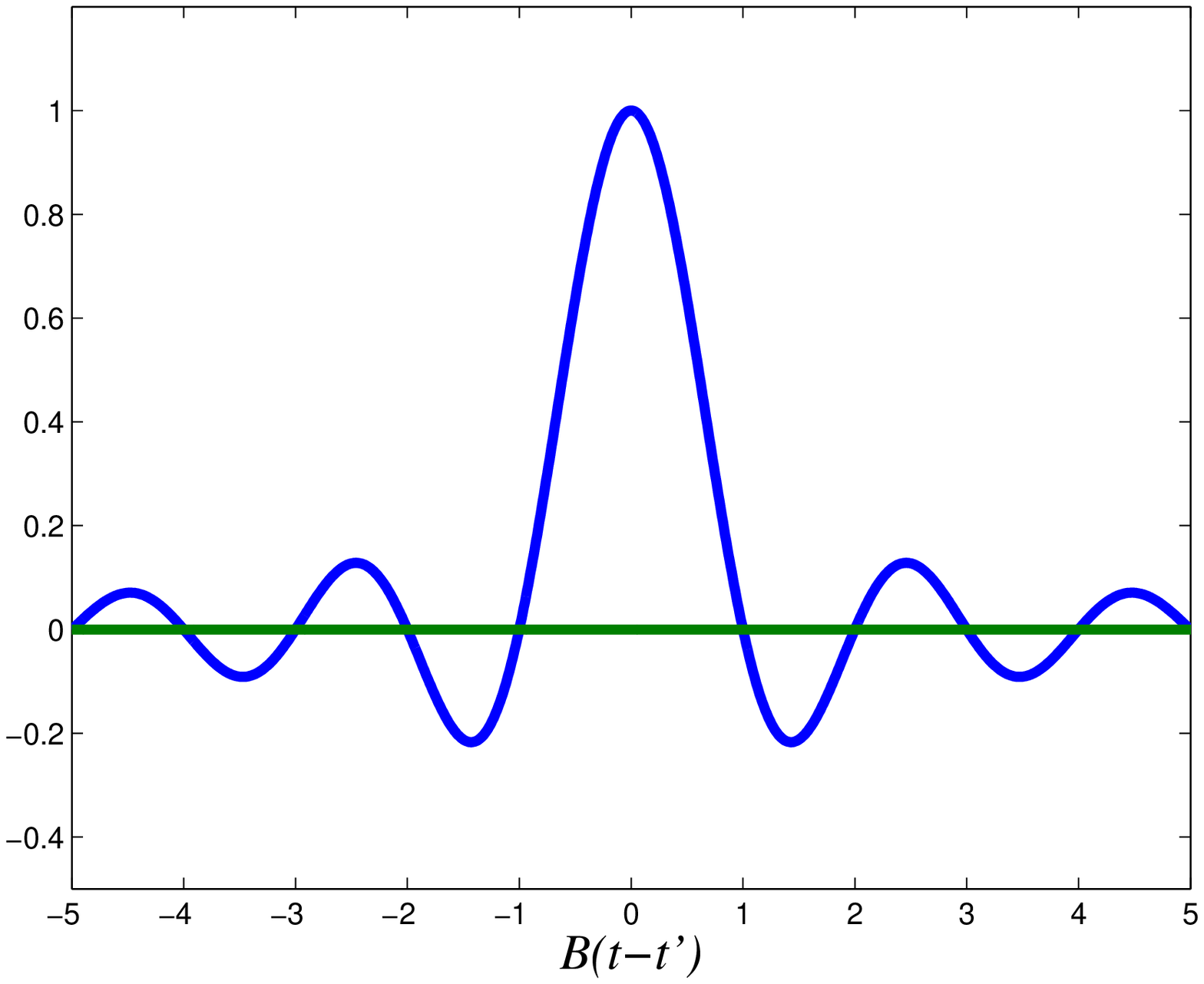}
\caption{(Color online). $\sinc B(t-t')$.}
\label{sinc}
\end{figure}

We make the important observation that the time-domain commutator
given by Eq.~(\ref{commutator}) goes to zero at discrete intervals, as
shown in Fig.~\ref{sinc}, when $t-t'$ is a nonzero-integer multiple of
$1/B$. If we discretize time into such intervals, that is, let
\begin{align}
t_j \equiv t_0 +j\delta t,
\end{align}
where $j$ is an integer and 
\begin{align}
\delta t \equiv \frac{1}{B},
\end{align}
the commutator at these
sampled times becomes
\begin{align}
[\hat{A}(t_j),\hat{A}^\dagger(t_k)] &= B\delta_{jk} =
\frac{\delta_{jk}}{\delta t}.
\end{align}
The sampled times have become orthogonal ``wave-packet'' modes, as
described qualitatively but not substantiated in
Refs.~\cite{yamamoto}. The operators can be renormalized to give the
conventional discrete-mode commutator,
\begin{align}
\hat{a}_j &\equiv \hat{A}(t_j)\sqrt{\delta t},
&
[\hat{a}_j,\hat{a}_k^\dagger] &= \delta_{jk},
\end{align}
and the continuous-time and continuous-frequency operators can be
reconstructed using the sampling theorem,
\begin{align}
\hat{A}(t) &= \sqrt{B}
\sum_{j=-\infty}^\infty\hat{a}_j \sinc B(t-t_j),
\label{sampling_time}\\
\hat{a}(f) &=\frac{1}{\sqrt{B}}
\sum_{j=-\infty}^\infty \hat{a}_j \exp(i2\pi f t_j)
\textrm{ for } |f|< \frac{B}{2}.
\end{align}
By virtue of the sampling theorem, the use of the discrete-time
operators on the bandlimited Hilbert space is completely
equivalent to the use of continuous-time or
continuous-frequency operators. 

The bandlimited Hilbert space can be spanned by photon-number states
in discrete wave-packet modes,
\begin{align}
\ket{n_j}_j&\equiv 
\frac{1}{\sqrt{n_j!}}\big(\hat{a}_j^{\dagger}\big)^{n_j}\ket{0}_j,
&
\ket{\bs{n}}
&\equiv \bigotimes_j \ket{n_j}_j,
\\
\hat{\bs{1}} &= \sum_{\bs{n}}
\ket{\bs{n}}\bra{\bs{n}},
\end{align}
where $n_j$ is the number of photons in the wave-packet mode at time
$t_j$, and for convenience we use boldface $\bs{n}$ to denote the vector
$\{\dots,n_j,n_{j+1},\dots\}$.  Appendix \ref{algebra} lists other
notations that we shall use to describe algebra in the discrete time
domain. For a pure state $\ket{\Psi}$, the photon-number
representation is
\begin{align}
C[\bs{n}]
&\equiv \avg{\bs{n}|\Psi},
&
\sum_{\bs{n}}|C[\bs{n}]|^2 &= 1.
\end{align}
Another way of spanning the Hilbert space is via nonorthogonal
multiphoton  time-measurement eigenstates \cite{mandel,shapiro_qso,fetter},
\begin{align}
\ket{\tau_1,\dots,\tau_N} &\equiv
\frac{1}{\sqrt{N!}}\hat{A}^\dagger(\tau_1)\dots\hat{A}^\dagger(\tau_N)\ket{0},
\\
\hat{\mathbf{1}} &= \sum_{N=0}^\infty\int d\tau_1\dots d\tau_N
\ket{\tau_1,\dots,\tau_N}\bra{\tau_1,\dots,\tau_N}.
\end{align}
The multiphoton wave function is then
\begin{align}
\psi_N(\tau_1,\dots,\tau_N)\equiv \avg{\tau_1,\dots,\tau_N|\Psi},
\\
\sum_{N=0}^\infty \int d\tau_1\dots d\tau_N
|\psi_N(\tau_1,\dots,\tau_N)|^2 = 1,
\end{align}
which is related to the photon-number representation by
\cite{fetter}
\begin{align}
\psi_N(\tau_1,\dots,\tau_N) &=
\sum_{\bs{n}, \sum_j n_j=N}
C[\bs{n}]
\Phi_{\bs{n}}(\tau_1,\dots,\tau_N),
\\
\Phi_{\bs{n}}(\tau_1,\dots,\tau_N) &\equiv
\bk{\frac{\prod_j n_j!}{N!\delta t^N}}^{1/2}\sum_{l=1}^{N!}
\nonumber\\&\quad 
\prod_j \prod_{r_j=\sum_k^{j-1} n_k+1}^{\sum_k^{j-1} n_k+n_j}
\sinc\bigg(\frac{\tau_{P_l(r_j)}-t_j}{\delta t}\bigg),
\end{align}
where $P_l(r_j)$ is an element of a permutation of
$\{r_1,\dots,r_N\}$ and the sum over $l$ is over all $N!$ possible
permutations so that $\Phi_{\bs{n}}$ is symmetric. The inverse
relation is
\begin{align}
 C[\bs{n}]&=
\bk{\frac{N!\delta t^N}{\prod_j n_j!}}^{1/2}
\nonumber\\&\quad\times
\psi_N(\dots,\underbrace{t_j,\dots,t_j}_{n_j{\rm terms}},
\underbrace{t_{j+1},\dots,t_{j+1}}_{n_{j+1}{\rm terms}},\dots),
\label{representations}
\end{align}
where $N = \sum_j n_j$.  For example, a coherent state can be defined
in the continuous time domain as
\begin{align}
\ket{\mathcal{A}(t)} &\equiv \exp\bigg[-\frac{\bar{N}}{2}+
\intall dt \mathcal{A}(t)\hat{A}^\dagger(t)\bigg]\ket{0},
\\
\hat{A}(t)\ket{\mathcal{A}(t)} &= \mathcal{A}(t)\ket{\mathcal{A}(t)},
\quad
\bar{N} = \intall dt |\mathcal{A}(t)|^2,
\end{align}
where $\mathcal{A}(t)$ is a bandlimited envelope function and $\bar{N}$ is
the average photon number. In terms of the wave-function
representation,
\begin{align}
\psi_N(\tau_1,\dots,\tau_N) &= 
\frac{\exp(-\bar{N}/2)}{\sqrt{N!}}\mathcal{A}(\tau_1)\dots 
\mathcal{A}(\tau_N).
\end{align}
Using Eq.~(\ref{representations}), we obtain the photon-number
representation for a coherent state,
\begin{align}
C[\bs{n}] &= \prod_j
\frac{\exp(-|\alpha_j|^2/2)}{\sqrt{n_j!}}\alpha_j^{n_j},
&
\alpha_j &= \mathcal{A}(t_j)\sqrt{\delta t},
\end{align}
which shows that any coherent state can be written as a collection of
independent coherent states in wave-packet modes.

\section{\label{canonical}Canonical temporal-phase measurements}
For each wave-packet mode at time $t_j$, we can define the
nonunitary Susskind-Glogower exponential-phase operator
\cite{susskind} as
\begin{align}
\sg_j &\equiv \frac{1}{\sqrt{\hat{a}_j\hat{a}_j^\dagger}}
\hat{a}_j =
\frac{1}{\sqrt{\hat{A}(t_j)\hat{A}^\dagger(t_j)}}
\hat{A}(t_j).
\end{align}
Since $\sg_j$ depends only on the field operators at time $t_j$, it
can be regarded as an instantaneous phase operator.  Despite the
nonunitary nature of $\sg_j$, its eigenstates
form a nonorthogonal basis of the Hilbert space,
\begin{align}
\ket{\bs\phi} &\equiv 
\sum_{\bs{n}}\exp (i\bs{n}\cdot\bs\phi)\ket{\bs{n}},
&
\bs\phi_0 &\le \bs\phi < \bs\phi_0 + 2\pi,
\\
\hat{\bs{1}} &= \int D\bs\phi
\ket{\bs\phi}\bra{\bs\phi},
&
D\bs\phi &\equiv \prod_j \frac{d\phi_j}{2\pi}.
\end{align}
The Susskind-Glogower eigenstates can thus be used to define a
temporal-phase positive operator-valued measure (POVM), also called a
probability operator measure (POM) \cite{shapiro,helstrom,holevo},
\begin{align}
\hat{\Pi}[\bs\phi]
&\equiv \ket{\bs\phi}
\bra{\bs\phi},
&
\int D\bs\phi \hat{\Pi}[\bs\phi]  &= \hat{\bs{1}},
\label{povm}
\end{align}
which corresponds to a measurement of instantantaneous optical phases
$\{\dots,\phi_j,\phi_{j+1},\dots\}$ at times
$\{\dots,t_j,t_{j+1},\dots\}$.  It can be shown, by generalizing the
single-mode treatment in Ref.~\cite{holevo}, that the POVM given by
Eq.~(\ref{povm}) corresponds to the optimal temporal-phase
measurements for any periodic cost function with nonnegative Fourier
coefficients. Following Leonhardt \textit{et al.}\ \cite{leonhardt},
we shall refer to the phase POVM measurement as the canonical phase
measurement.  In the single-mode case, an experimental canonical phase
measurement scheme has been proposed by Pregnell and Pegg
\cite{pregnell}. While generalization of their scheme to the time
domain is conceivable, it is beyond the scope of this paper to
investigate experimental canonical temporal-phase measurement schemes
in detail.

The canonical temporal-phase probability density is
\begin{align}
p[\bs\phi]
&\equiv
\trace\Big\{\hat{\rho}\hat{\Pi}[\bs\phi]\Big\},
&
\int D\bs\phi p[\bs\phi] &= 1.
\label{prob}
\end{align}
For a pure state, one can define a temporal-phase representation as
the discrete Fourier transform of the photon-number representation,
\begin{align}
\mathcal{C}[\bs\phi]
&\equiv \bra{\bs\phi}\Psi\rangle
= \sum_{\bs{n}}\exp(-i\bs{n}\cdot\bs\phi)
C[\bs{n}],
\label{fourier}
\end{align}
the magnitude squared of which gives the probability density
$p(\bs\phi)$.

One can also extend the Pegg-Barnett formalism to the time domain by
considering a Hilbert space spanned by finite-photon-number states
$\ket{\bs{n}}$ with $ \bs{0} \le \bs{n} \le \bs{s}$.
The commutator for $\hat{a}_j$ and $\hat{a}_j^\dagger$ becomes
\begin{align}
[\hat{a}_j,\hat{a}_k^\dagger] =
\delta_{jk}[1-(s_j+1)\ket{s_j}_j\bra{s_j}],
\end{align} although applying the commutator to physical states in the
$s_j \to \infty$ limit recovers the usual commutator
$[\hat{a}_j,\hat{a}_k^\dagger]_p = \delta_{jk}$. A unitary
Pegg-Barnett exponential-phase operator can be defined in the
photon-number basis as
\begin{align}
\exp(i\hat{\phi}_j) &\equiv 
\sum_{n_j=1}^{s_j}\ket{n_j-1}_j\bra{n_j}
+\exp[i(s_j+1)\phi_{0j}]\ket{s_j}_j\bra{0}.
\label{pb}
\end{align}
By taking the limit $s_j\to \infty$ at the end of calculations, the
Pegg-Barnett theory predicts the same phase statistics as the
canonical phase statistics governed by Eq.~(\ref{prob}), and the two
theories can be regarded as equivalent and complementary
\cite{shapiro}.

To define an instantaneous-frequency operator, consider the classical
definition
\begin{align}
F(t) &\equiv -\frac{1}{2\pi}\diff{}{t}\phi(t).
\end{align}
$\phi(t)$ can be multivalued or even undefined when the intensity
is zero, so to avoid this ambiguity we write the instantaneous
frequency in terms of $\exp[i\phi(t)]$,
\begin{align}
F(t) &= \frac{1}{4\pi i}\bigg\{\exp[i\phi(t)]\diff{}{t}
\exp[-i\phi(t)]-\textrm{c.c.}\bigg\}
\nonumber\\
&=\frac{1}{2\pi}\diff{}{t'}\sin[\phi(t)-\phi(t')]\Big|_{t'=t}
\end{align}
where c.c. denotes complex conjugate. In the quantum regime, we
can discretize $F(t)$ and express it exactly in terms of
the Pegg-Barnett operator,
\begin{align}
\hat{F}_j &= \frac{1}{2\pi\delta t}\sum_k d_{j-k}\sin(\hat\phi_j-\hat\phi_k),
\\
\sin(\hat\phi_j-\hat\phi_k) &= 
\frac{1}{2i}\bigg[\exp(i\hat\phi_j-i\hat\phi_k)-\textrm{H.c.}\bigg].
\end{align}
H.c. denotes Hermitian conjugate, and $d_{j-k}$ is called the
differentiator \cite{proakis}, the discrete
impulse response that corresponds to differentiation,
\begin{align}
d_{j-k} = \Big\{\begin{array}{cc}
(-1)^{j-k}/(j-k),& j \neq k,
\\
0, & j = k.
\end{array}
\label{differentiator}
\end{align}
$\hat{F}_j$ is Hermitian and well defined in the finite-photon-number
bandlimited Hilbert space, but given the difficulty of canonical phase
measurements in practice, it might be even more difficult to measure
the instantaneous-frequency operator, unless approximations are made.

\section{\label{map_estimation}Maximum A Posteriori Estimation}
It is difficult to perform canonical phase measurements, and
first-order field measurements, such as homodyne detection and
heterodyne detection, are often preferred in practice.  Statistics of
heterodyne detection and its variants \cite{noh} are governed by the Q
distribution \cite{leonhardt,yuen80,milburn,wagner}, which is broader
than the Wigner distribution that governs homodyne detection
\cite{yuen80,milburn}, so ideally one would like to use only homodyne
detection to measure the phase and instantaneous frequency. In this
section we would like to show how this can be done from the
perspective of estimation theory, which naturally leads to the use of
homodyne phase-locked loops to perform angle demodulation.

The Wigner distribution is defined as \cite{milburn}
\begin{align}
W[\bs{x},\bs{y}] &\equiv \int D\bs{b}
\exp(\bs{b}^*\cdot\bs{a}-\bs{b}\cdot\bs{a}^*)\chi[\bs{b}],
\\
D\bs{b} &\equiv \prod_j \frac{d^2b_j}{\pi^2},
\quad
\bs{a} \equiv \frac{1}{2}(\bs{x}+i\bs{y}),
\\
\chi[\bs{b}] &\equiv \trace\Big\{\hat{\rho}
\exp\Big(\bs{b}\cdot\hat{\bs{a}}^\dagger-
\bs{b}^*\cdot\hat{\bs{a}}\Big)\Big\},
\end{align}
which governs the statistics of operators that can be expressed in
Weyl-ordered quadrature operators \cite{mandel,milburn},
\begin{align}
\avg{f^{(W)}[\hat{\bs{x}},\hat{\bs{y}}]} 
&= \int D\bs{x} D\bs{y} f[\bs{x},\bs{y}]W[\bs{x},\bs{y}],
\\
\hat{\bs{x}} &\equiv  \hat{\bs a}+\hat{\bs a}^\dagger,
\quad
\hat{\bs{y}} \equiv -i(\hat{\bs a}-\hat{\bs a}^\dagger),
\\
D\bs{x} D\bs{y} &\equiv \prod_j dx_j dy_j,
\end{align}
where the superscript $(W)$ denotes Weyl ordering.
In the following we shall consider only quantum states that have
nonnegative Wigner distributions, and assume that measurements of
Weyl-ordered quadrature operators can always be realized.
The Wigner distribution is then a qualified classical probability
distribution that we can use in classical estimation theory.

If the range of phase modulation exceeds $2\pi$, one must unwrap the
measured phase to unambiguously decode the message contained
within. Phase unwrapping can be done only if one has a priori
information about the message, because otherwise one would have no
other way to distinguish $\phi_j$ from $\phi_j+2n\pi$, $n$ being an
arbitrary integer, in a measurement. To incorporate a priori
information in angle demodulation, we shall use maximum a
posteriori (MAP) estimation \cite{viterbi,vantrees}. Given an a priori
probability distribution of the message
$P[\bs{m}]=P(\dots,m_j,m_{j+1},\dots)$, MAP estimation seeks the
message that maximizes the a posteriori probability distribution
\begin{align}
P[\bs{m}|\bs{x},\bs{y}] &= \frac{W[\bs{x},\bs{y}|\bs{m}]P[\bs{m}]}
{P[\bs{x},\bs{y}]}
\end{align}
for a set of measurements in terms of $\bs{x}$ and $\bs{y}$, and is
asymptotically efficient. Taking the logarithm of the a posteriori
distribution and differentiating with respect to $\bs{m}$, we arrive
at the vectorial MAP equation, the solution of which gives the MAP
estimate $\est{\bs{m}}$,
\begin{align}
\Big(\nabla_{\bs m}\ln W[\bs{x},\bs{y}|\bs{m}] +
\nabla_{\bs{m}}\ln P[\bs{m}]\Big)_{\bs{m}=\est{\bs{m}}} &= \bs{0},
\label{map}
\end{align}
where the gradient operator $\nabla_{\bs m}$ is defined as
\begin{align}
\nabla_{\bs m} &\equiv \bigg\{\dots,\parti{}{m_j},\parti{}{m_{j+1}},\dots
\bigg\}.
\end{align}
The MAP equation (\ref{map}) can be significantly simplified if the
probability distributions are Gaussian. We therefore model the message
$\bs{m}$ as a zero-mean Gaussian random process,
\begin{align}
P[\bs{m}] &\propto \exp\Bigg(-\frac{1}{2}
\bs{m}\cdot\bs{K}_{\bs m}^{-1}\cdot\bs{m}\Bigg),
\label{message}
\\
\bs{K}_{\bs m} &\equiv \avg{\Delta \bs{m}\otimes\Delta\bs{m}},
\end{align}
where $\bs{K}_{\bs m}$ is the message covariance matrix.  In angle
modulation systems, the mean phase is a linear transformation of the
message,
\begin{align}
\bar{\bs\phi} &= \bs{H}\cdot\bs{m},
\end{align}
where $\bs{H}$ is a real impulse response. For example,
the impulse response for PM is
\begin{align}
\bs{H} = \beta \bs{I},
\end{align}
where $\beta$ is called the modulation index and $\bs{I}$ is the
identity matrix, and the impulse response for FM is
\begin{align}
H_{jk} = -2\pi \mathcal{F}\int_{-\infty}^{t_j} dt \sinc B(t-t_k),
\end{align}
where $\mathcal{F}$ is called the frequency deviation.  Although the Wigner
distribution is Gaussian for squeezed states
\cite{milburn}, we shall first study the simpler coherent states.

\subsection{Coherent states}
To derive the Wigner distribution for a coherent state, let us start with
that for the vacuum,
\begin{align}
W_0[\bs{x}_0,\bs{y}_0] \propto
\exp\bk{-\frac{1}{2}\bs{x}_0\cdot\bs{x}_0
-\frac{1}{2}\bs{y}_0\cdot\bs{y}_0}.
\label{vacuum}
\end{align}
A coherent state with mean phase $\bar{\bs\phi}$ and constant mean
amplitude $|\alpha|$ is obtained by displacing the vacuum along the
$\bs{x}_0$ quadrature followed by phase modulation. The resulting
Wigner distribution is
\begin{align}
W[\bs{x},\bs{y}|\bar{\bs\phi}]
&=
W_0\big[\bs{x}_0[\bs{x},\bs{y}],\bs{y}_0[\bs{x},\bs{y}]\big],
\\
\bs{x}_0 &= \bs x \cos\bar{\bs\phi}+\bs{y}\sin\bar{\bs\phi}
-2|\alpha|,
\\
\bs{y}_0 &=
-\bs{x}\sin\bar{\bs\phi}+\bs{y}\cos\bar{\bs\phi},
\end{align}
which can be used as the conditional distribution in the MAP equation
(\ref{map}).  After some algebra, the MAP equation becomes
\begin{align}
\est{\bs{m}} &= 2|\alpha|\bs{K}_{\bs m}\cdot\bs{H}^T\cdot \bs{p},
\label{general_map}
\end{align}
where $\bs{p}$ is a quadrature field that depends linearly on the
complex field $\bs{a}$ and nonlinearly on the estimated message
$\est{\bs m}$,
\begin{align}
\bs{p} &\equiv -i\big[\bs{a}\exp(-i\est{\bs\phi})-
\bs{a}^*\exp(i\est{\bs\phi})\big],
\label{p}
&
\est{\bs\phi} &\equiv \bs{H}\cdot\est{\bs m}.
\end{align}

\begin{figure}[htbp]
\includegraphics[width=0.45\textwidth]{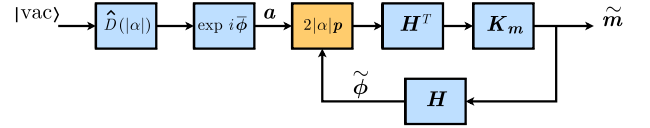}
\caption{(Color online). A block diagram that represents the MAP
  estimation process.  $\hat{D}(|\alpha|)$ denotes displacement of the
  vacuum along the $\bs{x}_0$ quadrature, and $\exp i\bar{\bs\phi}$
  denotes phase modulation.}
\label{coherent_map}
\end{figure}

The nonlinear MAP equation (\ref{general_map}) can be represented
conceptually by a block diagram, as shown in Fig.~\ref{coherent_map},
but cannot be solved analytically in general. Although the block
diagram seems to suggest that homodyne measurements of the quadrature
$\bs{p}$ in a feedback system can be used to produce the estimated
message $\est{\bs m}$, $\bs{p}$ depends recursively on
$\est{\bs\phi}$, and the impulse response matrix
$\bs{H}\cdot\bs{K}_{\bs m}\cdot\bs{H}^T$ that transforms $\bs{p}$ to
$\est{\bs\phi}$ is not necessarily causal.  Even if one attempts to
solve the MAP equation numerically, one needs the values of both
quadratures of $\bs{a}$ at all times, so one would have to perform
heterodyne detection and start with the Q distribution instead of the
Wigner distribution.

To solve Eq.~(\ref{general_map}) approximately, let us linearize it so
that the right-hand side of Eq.~(\ref{general_map}) depends linearly
on $\est{\bs{m}}$ and we can use well established linear estimation
techniques to perform MAP estimation. Writing the complex field
$\bs{a}$ in terms of $\bs{x}_0$ and $\bs{y}_0$ as
\begin{align}
\bs{a} &= \exp(i\bar{\bs\phi})\bigg(|\alpha|+
\frac{\bs{x}_0+i\bs{y}_0}{2}\bigg),
\end{align}
$\bs{p}$ can be rewritten as
\begin{align}
\bs{p} &=
2|\alpha|\sin(\bar{\bs\phi}-\est{\bs\phi})
+\bs{z}_0,
\end{align}
where we have defined another quadrature field $\bs{z}_0$ as
\begin{align}
\bs{z}_0 &\equiv \bs{x}_0\sin(\bar{\bs\phi}-\est{\bs\phi})
+\bs{y}_0\cos(\bar{\bs\phi}-\est{\bs\phi}).
\end{align}
Quantum noise enters the MAP estimation through the $\bs{z}_0$
quadrature. Because the vacuum Wigner distribution given by
Eq.~(\ref{vacuum}) is invariant with respect to rotations in phase
space, any quadrature in terms of $\bs{x}_0$ and $\bs{y}_0$ has the
same Gaussian statistics as $\bs{x}_0$ or $\bs{y}_0$ and its
statistics are independent of $\bar{\bs\phi}$ or $\est{\bs\phi}$.
$\bs{z}_0$ can therefore be regarded as an independent noise term.  To
linearize $\bs{p}$, we assume that the estimated phase is at all
times close to the mean phase of the field in the mean-square sense,
\begin{align}
\avg{(\bar{\bs\phi}-\est{\bs\phi})^2} \ll \bs{1},
\end{align}
so that we can make the first-order approximation
\begin{align}
\bs{p} &\approx 2|\alpha|(\bar{\bs\phi}-\est{\bs\phi})+\bs{z}_0.
\end{align}
A linearized MAP equation can thus be obtained,
\begin{align}
\est{\bs m} &\approx 
4|\alpha|^2\bs{K}_{\bs m}\cdot\bs{H}^T\cdot 
\bigg(\bar{\bs\phi}-\est{\bs\phi}+\frac{\bs{z}_0}{2|\alpha|}\bigg).
\label{linear_map}
\end{align}
In this linear regime the MAP estimation is efficient. Let us define
\begin{align}
\bs\phi &\equiv \bar{\bs\phi} + \frac{\bs{z}_0}{2|\alpha|},
&
\bs\epsilon &\equiv \bs\phi-\est{\bs\phi},
\end{align}
so that the linearized MAP equation (\ref{linear_map}) can be written
as a pair of equations,
\begin{align}
\est{\bs m} &= 
4|\alpha|^2 \bs{K}_{\bs m}\cdot\bs{H}^T
\cdot \bs\epsilon,
\label{map1}\\
\bs\phi-\bs{H}\cdot\est{\bs m} &= 
\bs\epsilon.
\label{map2}
\end{align}
Multiplying Eq.~(\ref{map1}) by $\bs{H}$ and adding
it to Eq.~(\ref{map2}),
\begin{align}
\bs\phi &= 
\big(4|\alpha|^2\bs{H}\cdot
\bs{K}_{\bs m}\cdot\bs{H}^T+\bs{I}\big)
\cdot\bs\epsilon.
\label{map3}
\end{align}
Comparing Eq.~(\ref{map3}) with Eq.~(\ref{map1}), we find that
$\est{\bs m}$ can be written as a linear transformation of $\bs\phi$,
\begin{align}
\est{\bs m} &= \bs{G}\cdot\bs\phi,
\label{linear_map2}
\end{align}
where $\bs{G}$ is the solution of the following
equation,
\begin{align}
\bs{G}\cdot\big(4|\alpha|^2\bs{H}\cdot
\bs{K}_{\bs m}\cdot\bs{H}^T+\bs{I}\big) &=
4|\alpha|^2\bs{K}_{\bs m}\cdot\bs{H}^T.
\label{optimum_filter}
\end{align}
To solve this equation, we assume that the message covariance matrix
$\bs{K}_{\bs m}$ is stationary, so that we can write
\begin{align}
(\bs{K}_{\bs{m}})_{jk} &= K_{\bs{m},j-k},
\end{align}
and define its power spectral density as
\begin{align}
S_{\bs m}(f) &\equiv \sum_j K_{\bs{m},j}\exp(i2\pi ft_j).
\end{align}
If $\bs{H}$ is time-invariant,
\begin{align}
H_{jk} &= H_{j-k}, &
H(f) &\equiv \sum_j H_j\exp(i2\pi ft_j),
\end{align}
we can solve Eq.~(\ref{optimum_filter}) by Fourier
transform and obtain a time-invariant solution for $\bs{G}$,
\begin{align}
G(f) &= \frac{4|\alpha|^2 S_{\bs m}(f)H^*(f)}
{4|\alpha|^2 S_{\bs m}(f)|H(f)|^2+1},
\\
G_{j-k} &= \frac{1}{B}\int_{-B/2}^{B/2}df G(f)\exp[-i2\pi f(t_j-t_k)].
\end{align}
$G_{j-k}$ is called the optimum filter in linear estimation theory
\cite{viterbi,vantrees}. In the next section we shall show how the
optimum filter can be implemented by a homodyne phase-locked loop.

\subsection{Phase-locked loops as angle demodulators}
\begin{figure}[htbp]
\includegraphics[width=0.45\textwidth]{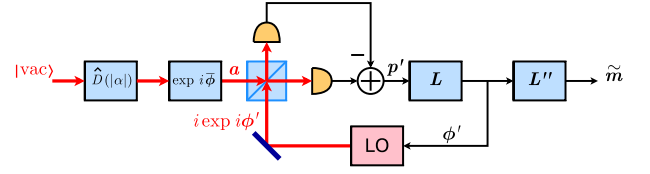}
\caption{(Color online). A homodyne phase-locked loop. LO denotes
  local oscillator.}
\label{pll}
\end{figure}

Consider the homodyne phase-locked loop shown in Fig.~\ref{pll}. The
output of the homodyne detection is given by
\begin{align}
\bs{p}' &=-i[\bs{a}\exp(-i\bs\phi')-\bs{a}^*\exp(i\bs\phi')]
\label{eta}\\
&=2|\alpha|\sin(\bar{\bs\phi}-\bs\phi')+\bs{z}',
\end{align}
where $\bs{z}'$ is another quadrature field defined as
\begin{align}
\bs{z}' &\equiv \bs{x}_0\sin(\bar{\bs\phi}-\bs\phi')
+\bs{y}_0\cos(\bar{\bs\phi}-\bs\phi'),
\label{z}
\end{align}
which, like $\bs{z}_0$, has Gaussian statistics independent of
$\bar{\bs\phi}$ and $\bs\phi'$.  If we again linearize $\bs{p}'$ by
making the assumption, to be justified later, that the
local-oscillator phase $\bs\phi'$ is at all times close to the mean
phase of the incoming field $\bar{\bs\phi}$,
\begin{align}
\avg{(\bar{\bs\phi}-\bs\phi')^2} \ll \bs{1},
\label{small_phase}
\end{align}
$\bs{p}'$ is given by
\begin{align}
\bs{p}' &\approx 2|\alpha|(\bs\phi-\bs\phi'),
&
\bs\phi &\equiv \bar{\bs\phi} + \frac{\bs{z}'}{2|\alpha|}.
\end{align}
$\bs\phi'$ is related to $\bs{p}'$ by the causal
loop filter $\bs{L}$,
\begin{align}
\bs\phi' &= \bs{L}\cdot\bs{p}'
\approx \bs{L}\cdot 2|\alpha|(\bs\phi-\bs\phi'),
\end{align}
and $\bs\phi'$ can be written in terms of
$\bs\phi$ and  a closed-loop filter $\bs{L}'$,
\begin{align}
\bs\phi' &\approx \bs{L}'\cdot \bs\phi,
\label{closed_loop}
\end{align}
where the closed-loop filter satisfy the equation
\begin{align}
(\bs{I}+2|\alpha|\bs{L})\cdot\bs{L}' &= 2|\alpha|\bs{L}.
\end{align}
To minimize $\avg{(\bar{\bs\phi}-\bs\phi')^2}$ and ensure that
Eq.~(\ref{small_phase}) holds, we demand $\bs\phi'$ to be the
minimum-mean-square-error estimate of $\bar{\bs\phi}$, using only past
and present values of $\bs{p}'$ to ensure causality. For stationary
$\bs\phi$, $\bar{\bs\phi}$, and $\bs{z}'$, $\bs{L}'$ is simply the
Wiener filter and can be solved by a well-known frequency-domain
technique \cite{viterbi,vantrees}, as briefly described below.
$L_{j-k}'$ is the solution of the discrete Wiener-Hopf equation,
\begin{align}
\sum_{k=0}^{\infty} L_{k}' U_{j-k} &= V_j,
\label{wiener_hopf}\\
\bs{U} &\equiv 4|\alpha|^2
\bs{H}\cdot\bs{K}_{\bs m}\cdot\bs{H}^T+\bs{I},
\\
\bs{V} &\equiv 4|\alpha|^2\bs{H}\cdot\bs{K}_{\bs m}\cdot\bs{H}^T.
\end{align}
The Fourier
transform of $U_{j-k}$ can be factored into the form
\begin{align}
U(f) &= X(f)X^*(f),
\end{align}
where $X(f)$ and $1/X(f)$ are causal filters, if $U(f)$ is a rational
spectral density.  Defining
\begin{align}
\bigg[\frac{V(f)}{X^*(f)}\bigg]_+ &\equiv 
\frac{1}{B}\int_{-B/2}^{B/2}df' \frac{V(f')}{X^*(f')}
\sum_{j=0}^\infty \exp[i2\pi (f-f')t_j],
\end{align}
the Fourier transform of $L'_{j-k}$ is given by
\begin{align}
L'(f) &= \frac{1}{X(f)}\bigg[\frac{V(f)}{X^*(f)}\bigg]_+,
\end{align}
and the loop filter is then
\begin{align}
L(f) &= \frac{L'(f)}{2|\alpha|[1-L'(f)]}.
\end{align}
It can be shown that the loop filter $\bs{L}$ is also causal given a
causal $\bs{L}'$ \cite{viterbi,vantrees}.
After passing through the phase-locked loop and the
post-loop filter $\bs{L}''$, the output estimated message becomes
\begin{align}
\est{\bs m} &\approx\bs{L}''\cdot\bs{L}'\cdot\bs\phi.
\end{align}
This equation is the same as the linearized MAP equation
(\ref{linear_map2}) if
\begin{align}
\bs{G} &= \bs{L}''\cdot\bs{L}'.
\end{align}
The post-loop filter in the frequency domain is thus given by
\begin{align}
L''(f) &= \frac{G(f)}{L'(f)}.
\end{align}
To make the post-loop filter realizable, we allow delay in the
estimated message, that is, we let
\begin{align}
\bar\phi_j &= \sum_k H_{j-k} m_{k+d},
&
\est\phi_j &=\sum_k H_{j-k}\est{m}_{k+d},
\end{align}
and $\bs{L}''$ can be made causal in the limit of infinite delay $d$
\cite{viterbi,vantrees}. In practice $d\delta t$ only needs to be a
few times larger than the correlation time of the message for
$\bs{L}''$ to be well approximated by causal filters.  Hence, the
homodyne phase-locked loop is able to realize the MAP estimation for
coherent states, provided that the approximation given by
Eq.~(\ref{small_phase}) is valid and the local-oscillator phase
follows closely the mean phase of the incoming field at all
times. This result is hardly surprising, because a coherent state can
be regarded as a classical signal with additive white Gaussian noise,
in which case it is well known that the homodyne phase-locked loop
realizes the optimum angle demodulator \cite{viterbi,vantrees}. Our
analysis so far closely mirrors that in classical estimation theory,
except that we have worked with discrete time for consistency with
previous sections. It is straightforward to take the continuous limit
of our analysis in this section, if one so desires, by redefining some
of the quantities and taking the limit $B\to \infty$.

\subsection{Squeezed states}
\begin{figure}[htbp]
\centerline{\includegraphics[width=0.48\textwidth]{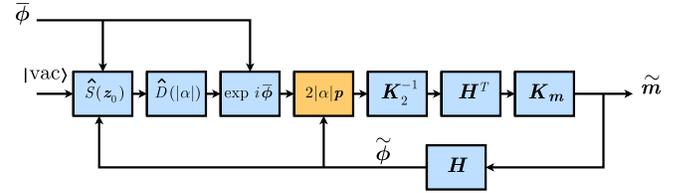}}
\caption{(Color online). MAP estimation for squeezed
  states. $\hat{S}(\bs{z}_0)$ denotes squeezing of the $\bs{z}_0$
  quadrature.}
\label{squeezed_map}
\end{figure}

The MAP estimation analysis shows that quantum noise enters the
estimation through the $\bs{z}_0$ quadrature, which depends on both
$\bs{x}_0$ and $\bs{y}_0$. If we squeeze the $\bs{z}_0$ quadrature of
the vacuum before displacement and phase modulation, the noise
entering the measurement of $\bs{p}$ can be reduced. The
squeezed-vacuum Wigner distribution is
\begin{align}
W_0[\bs{x}_0,\bs{y}_0] &\propto 
\exp\bigg(-\frac{1}{2}\bs\zeta_0\cdot\bs{K}_1^{-1}\cdot\bs\zeta_0-
\frac{1}{2}\bs{z}_0\cdot\bs{K}_2^{-1}\cdot\bs{z}_0
\bigg),
\\
\bs\zeta_0 &= \bs{x}_0\cos(\bar{\bs\phi}-\est{\bs\phi})
-\bs{y}_0\sin(\bar{\bs\phi}-\est{\bs\phi}),
\\
\bs{z}_0 &= \bs{x}_0\sin(\bar{\bs\phi}-\est{\bs\phi})
+\bs{y}_0\cos(\bar{\bs\phi}-\est{\bs\phi}),
\\
\bs{K}_1 &\equiv \avg{\bs\zeta_0\otimes\bs\zeta_0},
\quad
\bs{K}_2 \equiv \avg{\bs z_0\otimes \bs z_0}.
\end{align}
The MAP equation becomes
\begin{align}
\est{\bs m} &= 2|\alpha|\bs{K}_{\bs m}\cdot\bs{H}^T\cdot\bs{K}_2^{-1}
\cdot\bs{p},
\end{align}
as shown schematically in Fig.~\ref{squeezed_map}.  The linearized MAP
equation is the same as Eq.~(\ref{linear_map2}), but $\bs{G}$ now also
depends on $\bs{K}_2$,
\begin{align}
\bs{G}\cdot
\big(4|\alpha|^2\bs{H}\cdot
\bs{K}_{\bs m}\cdot\bs{H}^T+\bs{K}_2\big) &=
4|\alpha|^2\bs{K}_{\bs m}\cdot\bs{H}^T.
\label{squeezed_filter}
\end{align}

\begin{figure}[bhtp]
\centerline{\includegraphics[width=0.48\textwidth]{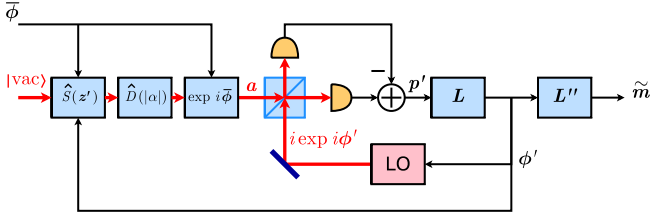}}
\caption{(Color online). A homodyne phase-locked loop that realizes
  the MAP estimation for squeezed states. $\hat{S}(\bs{z}')$ denotes
  squeezing of the $\bs{z}'$ quadrature.}
\label{squeezed_pll}
\end{figure}

A homodyne phase-locked loop, with redesigned loop and post-loop
filters $\bs{L}$ and $\bs{L}''$ according to the $\bs{G}$ given by
Eq.~(\ref{squeezed_filter}), can again be used to realize this MAP
estimation, as shown in Fig.~\ref{squeezed_pll}, although one must
feed the local-oscillator phase $\bs\phi'$ back to the quantum-state
preparation stage in order to squeeze the right quadrature of the
vacuum. The time delay of the feedback path must be much shorter than
the time scale at which the phase modulation varies.

If the feedback of $\bs\phi'$ and the $\bs{z}'$-quadrature squeezing
is not feasible in practice, another option is to use an ideal
phase-squeezed state and squeeze the $\bs{y}_0$ quadrature of the
vacuum before displacement and phase modulation. The squeezed-vacuum
Wigner distribution is
\begin{align}
W_0[\bs{x}_0,\bs{y}_0] &\propto 
\exp\bigg(-\frac{1}{2}\bs{x}_0\cdot\bs{K}_1^{-1}\cdot\bs{x}_0-
\frac{1}{2}\bs{y}_0\cdot\bs{K}_2^{-1}\cdot\bs{y}_0
\bigg).
\end{align}
The MAP equation becomes
\begin{align}
\est{\bs m} &= \bs{K}_{\bs m}\cdot\bs{H}^T\cdot\bs\eta,
\end{align}
where
\begin{align}
\bs\eta &\equiv \bs{q}(\bs{K}_2^{-1}\cdot\bs{p})+
\bs{p}\big(\bs{K}_1^{-1} \cdot(2|\alpha|-\bs{q})\big),
\\
\bs{q} &\equiv \bs{a}\exp(-i\est{\bs\phi})+\bs{a}^*\exp(i\est{\bs\phi}),
\\
\bs{p} &\equiv -i[\bs{a}\exp(-i\est{\bs\phi})-
\bs{a}^*\exp(i\est{\bs\phi})].
\end{align}
The MAP equation now depends quadratically on two quadrature fields
$\bs{q}$ and $\bs{p}$. Realizing this MAP estimation would require
some nontrivial second-order field measurements of $\bs\eta$ in a
Weyl-ordered operator form for the statistics to obey the
Wigner distribution.

\begin{figure}[htbp]
\includegraphics[width=0.48\textwidth]{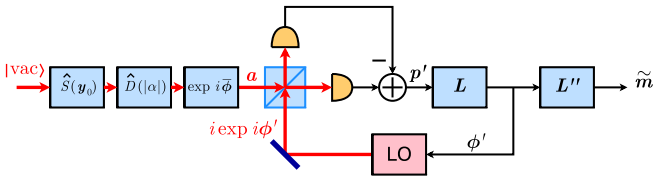}
\caption{(Color online). A homodyne phase-locked loop for
  phase-squeezed states.}
\label{squeezed_pll_no_feedback}
\end{figure}

If we insist on using the homodyne phase-locked loop for
phase-squeezed states, the noise term of the homodyne detection
$\bs{z}'$ in the first order of small
$\avg{(\bar{\bs\phi}-\bs\phi')^2}$ is given by
\begin{align}
\bs{z}' &\approx \bs{x}_0(\bar{\bs\phi}-\bs\phi')
+\bs{y}_0.
\end{align}
The $\bs{y}_0$ quadrature is squeezed, but the antisqueezed $\bs{x}_0$
quadrature also enters the phase-locked loop through the
phase-sensitive term $\bs{x}_0(\bar{\bs\phi}-\bs\phi')$, the magnitude
of which varies depending on the difference between the mean phase and
the local-oscillator phase. If the magnitude of
$\bs{x}_0(\bar{\bs\phi}-\bs\phi')$ is much smaller than the magnitude
of $\bs{y}_0$, however, we can assume that $\bs{z}'$ is approximately
$\bs{y}_0$,
\begin{align}
\bs{z}' &\approx \bs{y}_0,
\end{align}
and the homodyne phase-locked loop can still be used, as shown in
Fig.~\ref{squeezed_pll_no_feedback}, with the same filters as those
designed for the $\bs{G}$ given by Eq.~(\ref{squeezed_filter}).

Because $\bs{z}'$ passes through the loop filter $\bs{L}$ before
entering the system, the approximation $\bs{z}'\approx\bs{y}_0$
requires
\begin{align}
\avg{(\bar{\bs\phi}-\bs\phi')^2} \ll
\frac{\avg{(\bs{L}\cdot\bs{y}_0)^2}}{\avg{(\bs{L}\cdot\bs{x}_0)^2}}.
\label{squeezed_threshold}
\end{align}
If the constraint given by Eq.~(\ref{squeezed_threshold}) is violated,
$\bs{z}'$ cannot be regarded as a phase-insensitive noise term, and
the phase-locked loop is no longer an accurate realization of MAP
estimation for phase-squeezed states. It might still be possible to
heuristically adjust the parameters of the phase-locked loop to
compensate for the phase-sensitive noise and minimize the estimation
error, but Eq.~(\ref{squeezed_threshold}) sets a rough limit on the
right amount of squeezing, beyond which stronger squeezing is no
longer useful.

If one is able to perform canonical temporal-phase measurements, then
the homodyne detection in the phase-locked loop shown in
Fig.~\ref{squeezed_pll_no_feedback} can be replaced with measurements
of $\sin(\bs\phi-\bs\phi')$, where $\bs\phi$ is the canonical phase of
the incoming field $\bs{a}$. Canonical phase measurements are not
sensitive to the antisqueezed photon-number noise, so the squeezing
constraint given by Eq.~(\ref{squeezed_threshold}) is not needed,
although the constraint given by Eq.~(\ref{small_phase}) is still
required for the estimation to remain efficient in the linear regime.

The adaptive temporal-phase measurement scheme proposed by Berry and
Wiseman \cite{berry} is essentially a first-order homodyne
phase-locked loop, with an integrator as the loop filter and no
post-loop filter. While the first-order phase-locked loop is indeed
the optimal demodulator when the mean phase is a Wiener random process
\cite{vantrees}, it is clear from the preceding discussion that, at
least for stationary messages, one should use a more complicated loop
filter as well as a post-loop filter according to the a priori message
statistics to minimize the estimation error. In this paper we focus on
stationary messages and the use of MAP estimation to design angle
demodulators. For nonstationary messages, it is more appropriate to
use a state-variable approach and Kalman-Bucy filters
\cite{vantrees,kalman_bucy}, although it is beyond the scope of this
paper to study the latter case.

\section{\label{limits}Quantum limits to angle demodulation}
To calculate the mean-square error in the estimated message by MAP
estimation, consider the linearized MAP equation
\begin{align}
\est{\bs m} &= \bs{G}\cdot\bs\phi
=\bs{G}\cdot\bs{H}\cdot\bs{m}
+\bs{G}\cdot\frac{\bs{z}'}{2|\alpha|}.
\end{align}
The mean-square error is
\begin{align}
\avg{(\est{\bs m} - \bs m)^2}
&\approx \operatorname{Diag}\bigg\{
(\bs{G}\cdot\bs{H}-\bs{I})\cdot\bs{K}_{\bs m}\cdot(\bs{G}\cdot\bs{H}-\bs{I})^T
\nonumber\\&\quad
+\frac{1}{4|\alpha|^2}\bs{G}\cdot\bs{K}_{2}\cdot\bs{G}^T\bigg\},
\label{meansquare}
\end{align}
where $\operatorname{Diag}(\bs{A})_j = A_{jj}$ is the diagonal vector
component of $\bs{A}$. Equation (\ref{meansquare}) can be further
simplified in the Fourier domain to give
\begin{align}
\sigma^2 &\equiv \avg{(\est{m}_j- m_j)^2}
\nonumber\\
&\approx \frac{1}{B}\int_{-B/2}^{B/2}df \frac{S_{\bs m}(f)S_2(f)}
{4|\alpha|^2 S_{\bs m}(f)|H(f)|^2+S_2(f)},
\label{irreducible}
\end{align}
where $S_2(f)$ is the power spectral density of the $\bs{z}'$
quadrature.  This expression is well known in linear estimation theory
and called the irreducible error \cite{viterbi,vantrees}. Assuming
that the message has unit variance $\avg{m_j^2} = 1$ and a flat
spectral density with bandwidth $b$,
\begin{align}
S_{\bs m}(f) = \Big\{\begin{array}{cc}
B/b, & |f| < b/2,
\\
0, & |f| \ge b/2.
\end{array}
\label{msg_psd}
\end{align}
If $S_{2}(f)$ is also flat over the bandwidth $b$, the
mean-square error for PM becomes
\begin{align}
\sigma_{\textrm{PM}}^2 &= \frac{1}{\beta^2 \Lambda +1}
\approx 
\frac{1}
{\beta^2 \Lambda},
&
\beta^2\Lambda  &\gg 1,
\label{sigma_PM}
\end{align}
where we have defined the parameter $\Lambda$ as
\begin{align}
\Lambda &\equiv \frac{4|\alpha|^2S_{\bs m}(0)}{S_{2}(0)}.
\end{align}
The SNR is then
\begin{align}
{\rm SNR}_{\rm PM} &\equiv \frac{\avg{m_j^2}}{\sigma_{\rm PM}^2} 
\approx \beta^2 \Lambda.
\label{SNR_PM}
\end{align}
For FM, defining the equivalent modulation index as 
\begin{align}
\beta \equiv
\frac{2\mathcal{F}}{b},
\end{align}
the mean-square error is
\begin{align}
\sigma_{\textrm{FM}}^2 &= 
1-\sqrt{\beta^2\Lambda}
\tan^{-1}\frac{1}{\sqrt{\beta^2\Lambda}}
\approx \frac{1}{3\beta^2\Lambda},
\quad
\beta^2\Lambda \gg 1,
\label{sigma_FM}
\end{align}
and the SNR is
\begin{align}
{\rm SNR}_{\rm FM} &\approx 3\beta^2 \Lambda.
\label{SNR_FM}
\end{align}
The parameter $\Lambda$ depends on the ratio of the message spectral density
to the quadrature-noise spectral density.

\subsection{Standard and Heisenberg quantum limits}
For a coherent state with 
average optical power
\begin{align}
\mathcal{P} &\equiv hf_0 B\avg{\hat{a}^\dagger\hat{a}} = 
hf_0 B|\alpha|^2
\end{align}
and a quadrature-noise spectral density
\begin{align}
S_2(f) &= 1,
\end{align}
$\Lambda$ is
\begin{align}
\Lambda &= \frac{4\mathcal{P}}{hf_0 b} \equiv 4\mathcal{N},
&
\mathcal{N}\equiv \frac{\mathcal{P}}{hf_0 b},
\end{align}
where $\mathcal{N}$ is the average number of photons in a period of
$1/b$, and the $\sigma^2 \sim 1/\mathcal{N}$ dependence is an analog
of the usual standard quantum limit (SQL) in single-parameter
estimation \cite{glm}. We can then write the SNR at SQL as
\begin{align}
{\rm SNR}_{\rm PM} &\approx 
\frac{1}{3}{\rm SNR}_{\rm FM} \approx 4\beta^2 \mathcal{N}.
&
\textrm{(SQL)}
\label{sql}
\end{align}
This linear dependence on $\mathcal{N}$ may not hold for
nonbandlimited message spectral densities. For example, if we let the
message spectral density be Lorentzian,
\begin{align}
S_{\bs m}(f) &= \frac{B}{2\pi}\frac{b}{f^2+(b/2)^2},
\end{align}
the PM SNR for a coherent state in the limit of $B \gg b$ is
\begin{align}
{\rm SNR}_{\rm PM} &\approx
\sqrt{\frac{8\beta^2\mathcal{N}}{\pi}+1},
\end{align}
which scales with $\sqrt{\mathcal{N}}$ instead of $\mathcal{N}$.  A
similar $\sqrt{\mathcal{N}}$ scaling is also observed by Berry and
Wiseman for their adaptive temporal-phase measurement scheme with the
nonstationary Wiener process as the mean phase \cite{berry}. In the
following we consider only the bandlimited message spectral density
given by Eq.~(\ref{msg_psd}) for simplicity.

For squeezed states, it is shown in Appendix~\ref{squeeze} that
\begin{align}
S_2(f) = \exp(-2r) \textrm{ for } |f| < \frac{B_s}{2},
\end{align}
where $r$ is the squeeze parameter and $B_s$ is the squeeze
bandwidth. We should therefore make $B_s$ at least as large as the
message bandwidth $b$. The average power of a squeezed state is
\begin{align}
\mathcal{P} &= hf_0 B|\alpha|^2 + hf_0 B_s\sinh^2 r.
\end{align}
Assuming $B_s = b$,
\begin{align}
\Lambda &= 4(\mathcal{N}-\sinh^2 r)\exp(2r),
\label{squeezed_Lambda}
\end{align}
the SNRs become
\begin{align}
{\rm SNR}_{\rm PM} &\approx 
\frac{1}{3}{\rm SNR}_{\rm FM} \approx 4\beta^2 
(\mathcal{N} -\sinh^2 r)
\exp(2r).
\label{squeezed_SNR}
\end{align}
The optimal $\exp(2r)$ at which the SNRs in Eq.~(\ref{squeezed_SNR})
are maximum is given by
\begin{align}
\exp(2r) = 2\mathcal{N}+1,
\end{align}
and the maximum SNRs become
\begin{align}
{\rm SNR}_{\rm PM} &\approx 
\frac{1}{3}{\rm SNR}_{\rm FM} \approx 4\beta^2 
\mathcal{N}(\mathcal{N}+1).
&
(\textrm{Heisenberg})
\label{heisenberg}
\end{align}
The $\mathcal{N}^2$ scaling can be regarded as an analog of the
Heisenberg limit in quantum single-parameter estimation \cite{glm}.

\subsection{Threshold constraint}
It must be stressed that the preceding derivation of quantum limits
is accurate only in the linear regime, when the constraint given by
Eq.~(\ref{small_phase}) is satisfied. Beyond the linear regime, the
phase-locked-loop demodulator is no longer an efficient estimator, and
the SNR is expected to drop abruptly when the quantum noise is larger
than a certain threshold. Physically, the abrupt drop in SNR below
threshold is due to the periodic nature of $\bs{p}'$ with respect to
$\bar{\bs\phi}-\bs\phi'$, which can cause $\bs\phi'$ to differ from
$\bar{\bs\phi}$ by multiples of $2\pi$ when the noise is large. To
ensure that the estimation is efficient, Eq.~(\ref{small_phase}),
which we now call the threshold constraint in accordance with
classical theory, should be observed. Since we have assumed that the
$\bs{z}'$-quadrature spectral density is flat over the message bandwidth, an
expression for $\avg{(\bar\phi_j-\phi_j')^2}$ can be borrowed from
well known results for classical signals with additive white Gaussian
noise \cite{viterbi,vantrees}. For the flat message spectral density,
\begin{align}
\sigma_0^2 \equiv
\avg{(\bar\phi_j-\phi_j')^2} \approx \frac{1}{b\Lambda}\int_{-b/2}^{b/2}
df \ln \Bk{1+\Lambda |H(f)|^2}.
\end{align}
The threshold constraint for PM becomes
\begin{align}
\frac{1}{\Lambda}\ln (1+\beta^2\Lambda) \ll 1,
\label{threshold_PM}
\end{align}
while the constraint for FM is
\begin{align}
\frac{1}{\Lambda}
\bigg[\ln(1+\beta^2\Lambda)
+2\beta\sqrt{\Lambda}\tan^{-1}
\bigg(\frac{1}{\beta\sqrt{\Lambda}}\bigg)\bigg]
\ll 1.
\label{threshold_FM}
\end{align}
In practice, it has been determined numerically for classical signals
with additive white Gaussian noise that $\sigma_0^2 \le 0.25$ is
sufficient to satisfy the threshold constraint
\cite{viterbi,vantrees}, so a similar constraint on $\sigma_0^2$ can
be used for coherent states or $\bs{z}'$-quadrature-squeezed states.

\subsection{Squeezing constraint}
For phase-squeezed states and the homodyne phase-locked loop depicted
in Fig.~\ref{squeezed_pll_no_feedback}, one should also apply the
squeezing constraint (\ref{squeezed_threshold}) for the SNR to be
given by Eq.~(\ref{squeezed_SNR}). The order of magnitude of the
right-hand side of Eq.~(\ref{squeezed_threshold}) is calculated in
Appendix~\ref{squeeze} and is roughly $\exp(-4r)$.  The constraint
becomes
\begin{align}
\frac{\exp(4r)}{\Lambda}\ln (1+\beta^2\Lambda) \ll 1
\label{threshold_PM2}
\end{align}
for PM and
\begin{align}
\frac{\exp(4r)}{\Lambda}
\bigg[\ln(1+\beta^2\Lambda)
+2\beta\sqrt{\Lambda}\tan^{-1}
\bigg(\frac{1}{\beta\sqrt{\Lambda}}\bigg)\bigg]
\ll 1
\label{threshold_FM2}
\end{align}
for FM, where $\Lambda$ is given by Eq.~(\ref{squeezed_Lambda}). The
squeezing constraint is much more stringent than the threshold
constraint and more so for stronger squeezing, because the former
constraint requires the enhanced noise in the antisqueezed $\bs{x}_0$
quadrature to be negligible.

The squeezing constraint also prevents the homodyne phase-locked loop
with phase-squeezed states from reaching the Heisenberg limit given by
Eq.~(\ref{heisenberg}). Consider, for example, the left-hand side of
Eq.~(\ref{threshold_PM2}) at the Heisenberg limit,
\begin{align}
\frac{(2\mathcal{N}+1)^2}{4\mathcal{N}
(\mathcal{N}+1)}\ln (1+4\beta^2\mathcal{N}^2),
\end{align}
which is much smaller than 1 only when $4\beta^2\mathcal{N}^2 \ll 1$
and the exact SNR approaches its lowest possible value $1$, the a
priori SNR. If we let $\exp(2r)$ be a small parameter $\lambda$
times $\mathcal{N}$ instead,
\begin{align}
\exp(2r) &= \lambda\mathcal{N}, &\lambda &\ll 1,
&
\Lambda &\approx 4\lambda\mathcal{N}^2,
\end{align}
Eq.~(\ref{threshold_PM2}) becomes
\begin{align}
\lambda \ll 
\frac{4}{\ln(1+4\lambda\beta^2\mathcal{N}^2)}
\approx \frac{2}{\ln\mathcal{N}}
\end{align}
for $4\delta\beta^2\mathcal{N}^2 \gg 1$ and $\ln \mathcal{N} \gg \ln
(2\sqrt{\lambda}\beta)$. The SNR for PM is therefore limited by
\begin{align}
{\rm SNR}_{\rm PM} &\approx 4\lambda\beta^2\mathcal{N}^2
\ll \frac{8\beta^2\mathcal{N}^2}{\ln\mathcal{N}}
\end{align}
in the limit of large $\mathcal{N}$. The limit on the FM SNR
can be derived using a similar argument, and is given by
\begin{align}
\textrm{SNR}_{\rm FM} \ll \frac{24\beta^2\mathcal{N}^2}{\ln\mathcal{N}}
\end{align}
in the limit of large $\mathcal{N}$.  The
$\mathcal{N}^2/\ln\mathcal{N}$ scaling is analogous to the ultimate
limit to adaptive single-mode phase measurements
\cite{berry_limit}. 

\subsection{SNR enhancement by wideband angle modulation}
Apart from the quantum enhancement by squeezing, the SNRs above
threshold can be enhanced quadratically for a bandlimited message
spectral density simply by increasing the modulation index
$\beta$. This SNR enhancement is achieved at the expense of bandwidth
resource, as is well known in classical communication
\cite{viterbi,vantrees,couch}. The SNR enhancement effect for discrete
FM in the quantum regime has also been suggested by Yuen
\cite{yuen_book}. A good approximation of the optical signal bandwidth
$B$ when the phase or instantaneous frequency of a monochromatic wave
is modulated by a message with unit variance $\avg{m_j^2} = 1$,
bandwidth $b$, and modulation index $\beta$ is provided by Carson's
rule \cite{couch},
\begin{align}
B \approx (\beta+1)b.
\label{carson}
\end{align}
The modulation is described as narrowband when $\beta \ll 1$, wideband
when $\beta \sim 1$, and ultrawideband when $\beta \gg 1$. For
squeezed states, because they already have an inherent bandwidth $B_s$
before phase modulation, Carson's rule should be modified to read
\begin{align}
B \approx B_s + (\beta+1)b.
\end{align}
For a given $\Lambda$, $\beta$ is also limited by the threshold
constraint given by Eq.~(\ref{threshold_PM}), (\ref{threshold_FM}),
(\ref{threshold_PM2}), or (\ref{threshold_FM2}), since $\sigma_0^2$
increases logarithmically with $\beta$. Thus $\beta$ cannot be
increased indefinitely even if enough optical bandwidth is available.

\section{\label{sensing}Optical sensing of position and velocity}
\begin{figure}[htbp]
\includegraphics[width=0.35\textwidth]{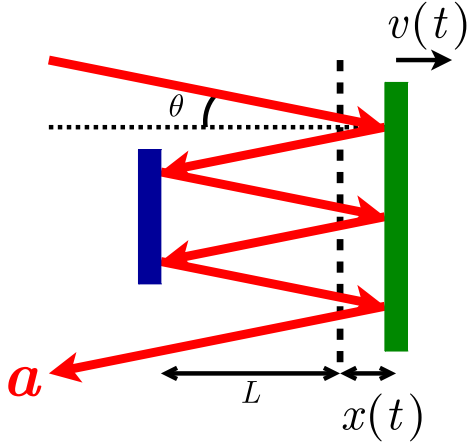}
\caption{(Color online). Multipass position and velocity sensor.}
\label{velocimeter}
\end{figure}

The preceding analysis can be applied directly to optical sensing of
position and velocity, if the phase modulation is due to reflection
off a classical moving target.  Consider the setup shown in
Fig.~\ref{velocimeter}.  A light beam, which can be a coherent state
or squeezed states, shines on a reflecting target and bounces back and
forth between the target and a perfectly reflecting mirror for $M$
times before leaving the apparatus. The phase modulation due to the
movement of the target is then estimated using canonical
temporal-phase measurements or a phase-locked loop. The mean phase of
the output light beam is given by
\begin{align}
\bar\phi(t) &=(2M\cos\theta)\frac{2\pi}{\lambda_0}  x(t)
\\
&=
(2M\cos\theta) \frac{2\pi f_0}{c}\int_{-\infty}^t 
dt'v(t').
\end{align}
The quantum limits derived in the preceding section are also
applicable here, if we assume that the target position and velocity
are stationary random processes and define the normalized PM and FM
parameters in terms of the sensing parameters as
\begin{align}
\beta &= 2M\cos\theta \frac{2\pi\sqrt{\avg{x^2(t)}}}{\lambda_0},
\\
m(t) &=\frac{x(t)}{\sqrt{\avg{x^2(t)}}},
\end{align}
for position sensing, and
\begin{align}
\mathcal{F} &=  (2 M\cos\theta)\frac{f_0\sqrt{\avg{v^2(t)}}}{c},
\\
\beta &=(2 M\cos\theta)\frac{2f_0\sqrt{\avg{v^2(t)}}}{bc},
\\
m(t) &= -\frac{v(t)}{\sqrt{\avg{v^2(t)}}}.
\end{align}
for velocity sensing. The power spectral densities of the position and
velocity are related to each other by
\begin{align}
S_x(f) &= \frac{1}{(2\pi f)^2}S_v(f),
\end{align}
so one should calculate the mean-square errors for position and
velocity estimations separately using Eq.~(\ref{irreducible}). It is
apparent from Eq.~(\ref{irreducible}) that increasing $\beta$ and
therefore $H(f)$ can at best reduce the mean-square errors of the
estimations quadratically. Even though the parameters $\avg{x^2(t)}$,
$\avg{v^2(t)}$, and $b$ are given for each target, it is still
possible to enhance the SNR by increasing $M$, the number of times the
target is interrogated. In practice the maximum achievable $M$ is
limited by the threshold constraint as well as other experimental
constraints. For instance, the total interrogation time of the light
beam with the target should be much shorter than the time scale at
which the position and the velocity changes, or in other words,
\begin{align}
\frac{2(M-1)L}{c\cos\theta} \ll \frac{1}{b}.
\end{align}
The target size, shape, and reflectivity also limit the maximum
achievable $M$ in practice.

The multipass setup may be regarded as a continuous-parameter
generalization of multipass single-parameter quantum estimation
schemes \cite{glm,higgins}. For single-parameter estimation, the
multipass SNR enhancement is achieved at the expense of time, whereas
for continuous-parameter estimation the enhancement effect utilizes
both time and bandwidth resources.  Squeezing becomes useful when such
resources are limited, and the experiment can be highly controlled to
eliminate decoherence, such as that caused by the imperfect
reflectivity of the target or other optical losses in the system.

\begin{figure}[htbp]
\includegraphics[width=0.4\textwidth]{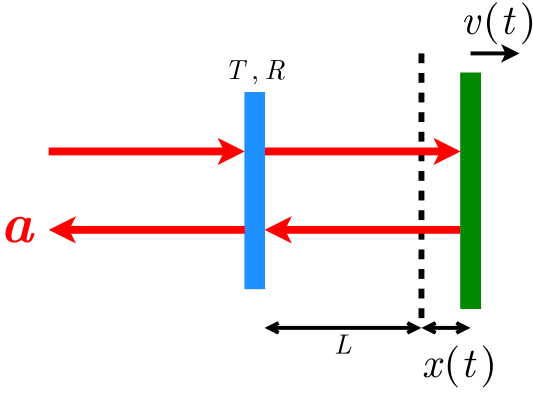}
\caption{(Color online). Fabry-P\'erot position and velocity sensor.}
\label{fb_velocimeter}
\end{figure}

To achieve a large effective $M$, one may also use a Fabry-P\'erot
setup, as shown in Fig.~\ref{fb_velocimeter}, similar to the
interferometric gravitational wave detector with Fabry-P\'erot arms
\cite{ligo}. $\cos\theta$ becomes $1$, and the effective $M$ at
resonance is
\begin{align}
M &= \frac{1+\sqrt{R}}{1-\sqrt{R}}.
\end{align}
Because the resonance condition is sensitive to the phase shift by the
target, the Fabry-P\'erot configuration is useful only in the
narrowband modulation regime $\beta \ll 1$.  Apart from this
restriction, the preceding discussion on the multipass configuration
applies to the Fabry-P\'erot setup as well.

\section{\label{fluid}Fluid velocity operator for nonrelativistic bosons}
We now apply the time-domain optical formalism described in
Sec.~\ref{quantize} and \ref{canonical} to nonrelativistic bosons in
three spatial dimensions. In momentum space, the boson annihilation
and creation operators obey the commutation relation
\begin{align}
[\hat{a}(\mathbf{k}),\hat{a}^\dagger(\mathbf{k}')]
&= \delta^3(\mathbf{k}-\mathbf{k}').
\end{align}
If we regard the nonrelativistic theory as an effective theory that
ceases to be accurate beyond a certain range of momenta, we can
explicitly impose a momentum cutoff to the Hilbert space, analogous to
the bandwidth limitation on optical fields in Sec.~\ref{quantize},
\begin{align}
|k_x|, |k_y|, |k_z| &< K,
\end{align}
so that we can discretize space as
\begin{align}
\mathbf{x}_{\mathbf{j}} &\equiv \mathbf{x}_0 + \mathbf{j}\delta x,
&
\mathbf{j} &\equiv j\mathbf{e}_x+k\mathbf{e}_y+l\mathbf{e}_z,
&
\delta x &\equiv \frac{\pi}{K}.
\end{align}
We can then define the space-domain operator as
\begin{align}
\hat{A}(\mathbf{x}) &=
\frac{1}{(2\pi)^{3/2}}\int_{-K}^{K} dk_x\int_{-K}^{K}dk_y\int_{-K}^{K}dk_z
\nonumber\\&\quad\times
\hat{a}(\mathbf{k})\exp(i\mathbf{k}\cdot\mathbf{x}),
\end{align}
and the discrete wave-packet-mode annihilation operator
as
\begin{align}
\hat{a}_{\mathbf{j}} &\equiv \hat{A}(\mathbf{x}_{\mathbf{j}})(\delta x)^{3/2},
\end{align}
with the commutator
\begin{align}
[\hat{a}_{\mathbf{j}},\hat{a}^\dagger_{\mathbf{j}'}] &= \delta_{\mathbf{j}\mathbf{j}'}.
\end{align}
The Hilbert space can be spanned by Fock states,
\begin{align}
\ket{n_{\mathbf{j}}}_{\mathbf{j}} &\equiv \sum_{n_{\mathbf{j}}=0}^\infty 
\frac{\big(\hat{a}_{\mathbf{j}}^{\dagger}\big)^{n_{\mathbf{j}}}}
{\sqrt{n_{\mathbf{j}}!}}\ket{0}_{\mathbf{j}},
&
\ket{\bs{n}} &\equiv \bigotimes_{\mathbf{j}} \ket{n_{\mathbf{j}}}_{\mathbf{j}}.
\end{align}
Unlike photons, the total number of massive bosons is conserved and
usually fixed, so it is physically justifiable to impose upper limits
to the boson numbers, so that the Hilbert space consists only
of finite-boson-number states,
\begin{align}
\hat{\bs{1}} &= \sum_{\bs{n}=\bs{0}}^{\bs{s}}\ket{\bs{n}}\bra{\bs{n}},
\end{align}
where $\bs{s} = \{\dots,s_{\mathbf{j}},\dots\}$ and for simplicity we let
$s_{\mathbf{j}}$ be the total number of bosons $N$.

With the finite-momentum and finite-boson-number Hilbert space, we can
now apply the Pegg-Barnett theory \cite{pegg} and define the unitary
exponential-phase operator as
\begin{align}
\exp(i\hat{\phi}_{\mathbf{j}}) 
&\equiv 
\sum_{n_{\mathbf{j}}=0}^{N}\ket{n_{\mathbf{j}}-1}_{\mathbf{j}}\bra{n_{\mathbf{j}}}
+
\exp[i(N+1)\phi_{0\mathbf{j}}]\ket{N}_{\mathbf{j}}\bra{0}.
\end{align}
Most importantly, the action of this operator on the vacuum state is
uniquely defined as
\begin{align}
\exp(i\hat{\phi}_{\mathbf{j}}) \ket{0}_{\mathbf{j}} &=
\exp[i(N+1)\phi_{0\mathbf{j}}]\ket{N}_{\mathbf{j}},
\label{lower_vacuum}
\end{align}
which makes the operator unitary and distinguishes it from the
nonunitary Susskind-Glogower operator.

The creation and annihilation operators can be rewritten as
\begin{align}
\hat{a}_{\mathbf{j}} &=\exp(i\hat{\phi}_{\mathbf{j}})\sqrt{\hat{n}_{\mathbf{j}}},
&
\hat{a}_{\mathbf{j}}^\dagger &= \sqrt{\hat{n}_{\mathbf{j}}}\exp(-i\hat{\phi}_{\mathbf{j}}).
\end{align}
Since the Hamiltonian is a function of $\hat{a}_{\mathbf{j}}$ and
$\hat{a}_{\mathbf{j}}^\dagger$, the unphysical jump from the vacuum state to
the maximum-number state indicated by Eq.~(\ref{lower_vacuum}) due to
the action of the Pegg-Barnett operator cannot occur directly in the
dynamics.

The commutation relation between the Pegg-Barnett operator and the
number operator $\hat{n}_{\mathbf{j}}\equiv
\hat{a}_{\mathbf{j}}^\dagger\hat{a}_{\mathbf{j}}$ is
\begin{align}
[\exp(i\hat{\phi}_{\mathbf{j}}),\hat{n}_{\mathbf{j}}] &=
[1-(N+1)\ket{N}_{\mathbf{j}}\bra{N}]
\exp(i\hat{\phi}_{\mathbf{j}}).
\label{pb_commutator}
\end{align}
To define a fluid velocity operator, consider the semiclassical definition
of a superfluid velocity field \cite{anderson},
\begin{align}
\mathbf{v}(\mathbf{x}) &\equiv \frac{\hbar}{m}\nabla_{\mathbf{x}}\phi(\mathbf{x}),
\label{gradient_phase}
\end{align}
where $m$ is the mass of each boson. Because $\phi(\mathbf{x})$ can be
a multivalued function and is undefined where the fluid density is
zero, Eq.~(\ref{gradient_phase}) is ill-defined even in the
semiclassical regime. An alternative is to consult the definition of
the current density \cite{london,dashen},
\begin{align}
\mathbf{J}(\mathbf{x}) &\equiv \frac{\hbar}{2im}
\Big[A^*(\mathbf{x})\nabla_{\mathbf{x}} A(\mathbf{x})
-A(\mathbf{x})\nabla_{\mathbf{x}} A^*(\mathbf{x})\Big]
\end{align}
and define the analogous fluid velocity in terms of the
exponential-phase function,
\begin{align}
\mathbf{v}(\mathbf{x}) &\equiv \frac{\hbar}{2im}
\Big\{\exp[-i\phi(\mathbf{x})]\nabla_{\mathbf{x}} 
\exp[i\phi(\mathbf{x})]
\nonumber\\&\quad
-\exp[i\phi(\mathbf{x})]\nabla_{\mathbf{x}}\exp[-i\phi(\mathbf{x})] \Big\}
\\
&=\frac{\hbar}{m}\nabla_{\mathbf{x}'} \sin[\phi(\mathbf{x}')-\phi(\mathbf{x})]
\Big|_{\mathbf{x}'=\mathbf{x}},
\end{align}
which is equivalent to Eq.~(\ref{gradient_phase}) where
$\phi(\mathbf{x})$ is continuous, and the sine function is always
single-valued where $\phi(\mathbf{x})$ is defined.  But we have not
yet solved the problem of defining $\phi(\mathbf{x})$ where the
density is zero. This can be done by regarding $\phi(\mathbf{x})$ as a
random function, or in the quantum regime, an operator. In the
discrete space domain, we can define the fluid velocity operator as
\begin{align}
\hat{\mathbf{v}}_{\mathbf{j}} &\equiv \sum_{\mathbf{j}'}
\mathbf{D}_{\mathbf{j}-\mathbf{j}'}
\sin(\hat{\phi}_{\mathbf{j}'}-\hat{\phi}_{\mathbf{j}}),
\end{align}
where
\begin{align}
\sin(\hat{\phi}_{\mathbf{j}'}-\hat{\phi}_{\mathbf{j}})
&\equiv \frac{1}{2i}
\Big[\exp(i\hat{\phi}_{\mathbf{j}'}-i\hat\phi_{\mathbf{j}})-
\textrm{H.c.}\Big].
\end{align}
$\mathbf{D}_{\mathbf{j}-\mathbf{j}'}$ is the
discrete impulse response that corresponds to the gradient operator,
\begin{align}
\mathbf{D}_{\mathbf{j}-\mathbf{j}'} &\equiv 
\frac{1}{\delta x}
(d_{j-j'}\delta_{kk'}\delta_{ll'}\mathbf{e}_x+
d_{k-k'}\delta_{jj'}\delta_{ll'}
\mathbf{e}_y
\nonumber\\&\quad+
d_{l-l'}\delta_{jj'}\delta_{kk'}\mathbf{e}_z),
\label{discrete_grad}
\end{align}
where $d_{j-j'}$ is the differentiator given by
Eq.~(\ref{differentiator}).  Neglecting the
$\ket{N}_{\mathbf{j}}\bra{N}$ term in the commutator in
Eq.~(\ref{pb_commutator}), which is important only in the highly
unlikely event that all bosons are in the same spatial wave-packet
mode, we arrive at the following velocity-number commutation relation,
\begin{align}
[\hat{v}_{\mathbf{j}},\hat{n}_{\mathbf{j}'}]
&\approx -i\frac{\hbar}{m}\mathbf{D}_{\mathbf{j}-\mathbf{j}'}
\cos(\hat\phi_{\mathbf{j}'}-\hat\phi_{\mathbf{j}}).
\label{discrete_landau}
\end{align}
In the limit of $\delta x \to 0$,
\begin{align}
\hat{\mathbf{v}}_{\mathbf{j}} &\to\hat{\mathbf{v}}(\mathbf{x}),
\\
\hat{n}_{\mathbf{j}} &\to \delta x^3\hat{\rho}(\mathbf{x}),
\\
\mathbf{D}_{\mathbf{j}-\mathbf{j}'} &\to \delta x^3
\nabla_{\mathbf{x}}\delta^3(\mathbf{x}-\mathbf{x}'),
\label{grad_limit}\\
\cos(\hat\phi_{\mathbf{j}'}-\hat\phi_{\mathbf{j}}) &\to 
\cos[\hat\phi(\mathbf{x}')-\hat\phi(\mathbf{x})],
\end{align}
we have
\begin{align}
[\hat{\mathbf{v}}(\mathbf{x}),\hat\rho(\mathbf{x}')]
&\approx -i\frac{\hbar}{m}
\nabla_{\mathbf{x}}\delta^3(\mathbf{x}-\mathbf{x}')
\cos[\hat\phi(\mathbf{x}')-\hat\phi(\mathbf{x})]
\nonumber\\
&\approx -i\frac{\hbar}{m}\nabla_{\mathbf{x}}\delta^3(\mathbf{x}-\mathbf{x}'),
\end{align}
the last expression of which agrees with the commutation relation
proposed by Landau \cite{landau}.  Thus, we have shown that it is
possible to rigorously define Landau's fluid velocity operator, if we
start with the physically reasonable assumptions of finite momentum
and finite boson number and take the continuous limit at the end of a
calculation. If the continuous limit is taken, one does not have
to use the $\mathbf{D}_{\mathbf{j}-\mathbf{j}'}$ defined in
Eq.~(\ref{discrete_grad}), and any
$\mathbf{D}_{\mathbf{j}-\mathbf{j}'}$ that possesses the continuous
limit given by Eq.~(\ref{grad_limit}) will suffice.

The fluid velocity is a physical quantity that can be measured
optically. The bosonic fluid can act as a moving dielectric, in which
a propagating light beam can acquire a phase shift proportional to the
fluid velocity in the first order, due to the Fresnel drag effect
\cite{leonhardt_moving}. Bosons may also reflect or scatter light, and
the fluid velocity will shift the instantaneous frequency of the
reflected or scattered light due to the Doppler effect.  Thus, the
fluid velocity and the optical instantaneous frequency are closely
related physical quantities, and, as we have shown in this paper, can
be described by the same theoretical formalism in the quantum regime.

\section{Conclusion}
In conclusion, we have proposed a quantum theory of optical phase and
instantaneous frequency in the time domain. In the formalism we have
introduced an explicit optical bandwidth cutoff, in order to reflect
our experimental limitations, satisfy the slowly-varying envelope
approximation, and, most importantly, discretize time domain into
discrete modes, the phases of which we know how to define and measure
in principle. Guided by insights from classical estimation theory, we
have suggested the use of homodyne phase-locked loops to perform angle
demodulation, and the quantum limits are derived. We have also shown
how the SNR enhancement effect of wideband angle modulation can be
applied to optical sensing of position and velocity. Given the recent
experimental advances in quantum-enhanced measurements
\cite{glm,higgins,goda} and the maturity of optical phase-locked loop
technology \cite{ferrero}, we expect our theoretical predictions to be
experimentally realizable with current technology and relevant to
future communication and sensing applications. Finally, we have
applied the optical formalism to nonrelativistic bosons and shown how
Landau's fluid velocity operator can be defined rigorously, thus
resolving a long-standing issue in quantum hydrodynamics. A more
rigorous formulation of the current-algebra approach to quantum
hydrodynamics based on our theory can be envisaged.

\section*{Acknowledgments}
This work is financially supported by the W.\ M.\ Keck Foundation
Center for Extreme Quantum Information Theory.

\appendix
\section{\label{algebra}Algebra in the discrete time domain}
For convenience we adopt the following simplified notations to
describe algebra in the discrete time domain.  Boldface lowercase
letters denote vectors in the discrete time domain,
\begin{align}
\bs{a} \equiv \{\dots,a_j,a_{j+1},\dots\}.
\end{align}
Multiplication and division of components of two vectors are 
written implicitly,
\begin{align}
\bs{a}\bs{b} &\equiv \{\dots,a_jb_j,a_{j+1}b_{j+1},\dots\},
\label{implicit_prod}\\
\frac{\bs{a}}{\bs{b}} &\equiv \bigg\{\dots,\frac{a_j}{b_j},
\frac{a_{j+1}}{b_{j+1}},\dots\bigg\}.
\label{implicit_div}
\end{align}
A vector of functions of each component of $\bs{a}$ is written
as a function of the vector,
\begin{align}
f(\bs{a}) &\equiv \{\dots,f(a_j),f(a_{j+1}),\dots\},
\end{align}
while a function of all components is written
with square brackets,
\begin{align}
f[\bs{a}] \equiv f(\dots,a_j,a_{j+1},\dots).
\end{align}
Dot product of two vectors is defined as
\begin{align}
\bs{a}\cdot\bs{b}\equiv \sum_j a_jb_j.
\end{align}
The tensor product of two vectors is
\begin{align}
(\bs{a}\otimes\bs{b})_{jk} = a_jb_k.
\end{align}
A matrix $A_{jk}$ is written as boldface and uppercase
$\bs{A}$.  The transpose of a matrix is defined as
\begin{align}
(\bs{A}^T)_{jk} \equiv A_{kj}.
\end{align}
Dot products of a matrix and a vector are
\begin{align}
(\bs{A}\cdot\bs{b})_j &\equiv \sum_k A_{jk}b_k,
&
(\bs{b}\cdot\bs{A})_j &\equiv \sum_k b_k A_{kj}.
\end{align}
Multiplication of two matrices is
\begin{align}
(\bs{A}\cdot\bs{B})_{jk} \equiv \sum_l A_{jl}B_{lk}.
\end{align}
The inverse of a matrix, if it exists, is $\bs{A}^{-1}$,
\begin{align}
\bs{A}\cdot\bs{A}^{-1} = \bs{A}^{-1}\cdot\bs{A} = \bs{I},
\end{align}
where $I_{jk} = \delta_{jk}$ is the identity matrix.

\section{\label{squeeze}Broadband squeezed vacuum}
We shall study the statistics of the squeezed vacuum
\cite{mandel,caves} in the interaction picture. For simplicity we
assume that the squeezing spectrum is uniform and has a bandwidth of
$B_s$. Starting with the vacuum quantum state, the output annihilation
operator in the frequency domain of a bandlimited squeezed vacuum can
be written as
\begin{align}
\hat{a}(f) &= 
\Big\{\begin{array}{cc}
\mu \hat{a}_{\textrm{vac}}(f) + \nu \hat{a}_{\textrm{vac}}^\dagger(-f),
& |f| < B_s/2,
\\
\hat{a}_{\textrm{vac}}(f) , & |f| \ge B_s/2,
\end{array}
\label{sq_vacuum}
\end{align}
where $\mu$ and $\nu$ are parametric gain parameters that satisfy
\begin{align}
|\mu|^2 - |\nu|^2 = 1,
\end{align}
and $\hat{a}_{\textrm{vac}}(f)$ is the annihilation operator with
vacuum state statistics. Performing the inverse Fourier transform
on Eq.~(\ref{sq_vacuum}),
\begin{align}
\hat{a}_j &\equiv \frac{1}{\sqrt{B}}\int_{-B/2}^{B/2} df \hat{b}(f)
\exp(-i2\pi f t_j),
\\
\hat{a}_{\textrm{vac},j} &\equiv
\frac{1}{\sqrt{B}}\int_{-B/2}^{B/2} df \hat{a}_{\textrm{vac}}(f)
\exp(-i2\pi f t_j),
\end{align}
the discrete-time-domain operator for a squeezed vacuum is
\begin{align}
\hat{\bs{a}} &= \hat{\bs{a}}_{\textrm{vac}}+
\bs{\Gamma}\cdot\Big[(\mu-1)\hat{\bs{a}}_{\textrm{vac}}+
\nu\hat{\bs{a}}_{\textrm{vac}}^\dagger\Big],
\end{align}
where $\Gamma_{jk} = \Gamma_{j-k}$ is a low-pass filter defined as
\begin{align}
\Gamma_{j-k} &\equiv \frac{1}{B}\int_{-B_s/2}^{B_s/2}df \exp[-i2\pi f (t_j-t_k)].
\end{align}
We define the quadrature operators as
\begin{align}
\hat{\bs{x}}_0 &\equiv \hat{\bs{a}}+\hat{\bs{a}}^\dagger,
&
\hat{\bs{y}}_0 &\equiv -i(\hat{\bs{a}}-\hat{\bs{a}}^\dagger).
\end{align}
The vacuum-state operators have the following statistics,
\begin{align}
\avg{\hat{\bs{a}}_{\textrm{vac}}}
&= \avg{\hat{\bs{a}}_{\textrm{vac}}^\dagger} = 0,
\\
\avg{\hat{\bs{a}}_{\textrm{vac}}^\dagger\otimes \hat{\bs{a}}_{\textrm{vac}}} &= 0,
\quad
\avg{\hat{\bs{a}}_{\textrm{vac}}
\otimes\hat{\bs{a}}_{\textrm{vac}}^\dagger} = \bs{I},
\end{align}
which lead to the following covariance matrices,
\begin{align}
\bs{K}_{1} 
&\equiv  \avg{\hat{\bs{x}}_0\otimes\hat{\bs{x}}_0}
=\bs{I}-\big(1-\abs{\mu+\nu^*}^2\big)\bs\Gamma,
\\
\bs{K}_{2} 
&\equiv \avg{\hat{\bs{y}}_0\otimes\hat{\bs{y}}_0}
= \bs{I}-\big(1-\abs{\mu-\nu^*}^2\big)\bs\Gamma.
\end{align}
To produce squeezing in the $\bs{y}_0$ quadrature, both $\mu$ and $\nu$
should be real and positive. Writing $\mu$ and $\nu$ in terms of the
squeeze parameter $r$,
\begin{align}
\mu &= \cosh r, & \nu &= \sinh r,
\end{align}
The covariance matrices become
\begin{align}
\bs{K}_{1} &= 
\bs{I}-\big[1-\exp(2r)\big]\bs\Gamma,
\\
\bs{K}_{2} 
&= \bs{I}-\big[1-\exp(-2r)\big]\bs\Gamma.
\end{align}
The quadrature power spectral densities are
\begin{align}
S_1(f) 
&=\Big\{\begin{array}{cc}
\exp(2r), & |f| < B_s/2,
\\
1, & |f| \ge B_s/2,
\end{array}
\\
S_2(f) 
&=\Big\{\begin{array}{cc}
\exp(-2r), & |f| < B_s/2,
\\
1, & |f| \ge B_s/2.
\end{array}
\end{align}
If $\hat{\bs{x}}_0$ and $\hat{\bs{y}}_0$ pass through a low-pass filter
$\bs{L}$ with bandwidth $b<B_s$, their variances become
\begin{align}
\avg{(\bs{L}\cdot\hat{\bs{x}}_0)^2} &= \frac{b}{B}\exp(2r),
\\
\avg{(\bs{L}\cdot\hat{\bs{y}}_0)^2} &= \frac{b}{B}\exp(-2r),
\end{align}
and their ratio is
\begin{align}
\frac{\avg{(\bs{L}\cdot\hat{\bs{y}}_0)^2}}
{\avg{(\bs{L}\cdot\hat{\bs{x}}_0)^2}} = \exp(-4r),
\end{align}
which provides an estimate of the magnitude of the right-hand side of
the threshold constraint given by Eq.~(\ref{squeezed_threshold}), if
we regard $\bs{L}$ as a bandpass filter with bandwidth $b$.

\end{document}